\documentclass{IEEEcsmag}

\usepackage[colorlinks,urlcolor=blue,linkcolor=blue,citecolor=blue]{hyperref}

\usepackage{upmath}
\usepackage{booktabs}
\usepackage{multirow}
\usepackage{lipsum}
\usepackage{threeparttable}
\usepackage[normalem]{ulem}
\usepackage{soul}
\usepackage{color}
\usepackage{xcolor}
\useunder{\uline}{\ul}{}

\jvol{XX}
\jnum{XX}
\paper{8}
\jmonth{September}
\jname{IEEE Software}
\pubyear{2022}

\begin{document}

\sptitle{Department: Head}
\editor{Editor: Name, xxxx@email}

\title{Co-designing for a Hybrid Workplace Experience in Software Development}

\author{Zhendong Wang$^{\dagger\ast}$, Yi-Hung Chou$^{\dagger\ast}$, Kayla Fathi$^{\ddagger}$, Tobias Schimmer$^{\ddagger}$, Peter Colligan$^{\ddagger}$, David Redmiles$^{\dagger}$, and Rafael Prikladnicki$^\mathsection$}
\affil{$^\dagger$University of California, Irvine, USA\\ $^{\ddagger}$SAP Labs, USA\\
$^\mathsection$Pontifical Catholic University of Rio Grande do Sul, Brazil}

\markboth{Department Head}{Co-design Hybrid Work}

\begin{abstract}
With increasing demands for flexible work models, many IT organizations have adapted to hybrid work that promises enhanced team productivity as well as work satisfaction. To achieve productive engineering practice, collaborative product innovation, and effective mentorship in the ensuing hybrid work, we introduce a workshop approach on co-designing for a hybrid workplace experience and provide implications for continuously improving collaborative software development at scale.
\end{abstract}

\maketitle

\chapterinitial{As} COVID-19 indiscriminately circled the globe, millions of workers had to adapt to a fully remote work environment. {\let\thefootnote\relax\footnotetext{*: This work was conducted while Zhendong and Yi-Hung interning at SAP.}}
Since this massive and widespread shift in work conditions, software practitioners have been re-evaluating their work preferences. According to increasing reports of shifts in work preferences, a plurality of information workers prefer a new normal of hybrid arrangements, which can yield improved flexibility and productivity concurrently.

The present article describes our practice of designing a hybrid workplace experience at one branch office of SAP Labs. This office tried to adopt a version of hybrid work: a work model combining in-office and remote work schedules. This model intends to improve collaboration quality with in-person sessions while also providing flexible and focused work time~\cite{ford2021tale}.

We took on a \textit{co-design} approach to address issues of transitioning towards a hybrid work model. We deliberately included various roles of software practitioners into the process of designing hybrid workplace experiences: prior studies suggest that creativity can be a wellspring of inspiration for everyone, especially when it comes to experiences that are closely meaningful to them~\cite{sanders2008co}. 
Besides, we argue that following a bottom-up decision-making process improved developer satisfaction during the transition, and potentially enhanced the future hybrid experience.
Thus, we leveraged co-design workshops to devise and improve software developers' hybrid workplace, and also incorporated the site's design thinking practice for inspiring participants' creativity.

The main question behind this work is:

\textit{How to design for and continuously improve a workplace experience with software development teams while transitioning to a hybrid work arrangement?}

To answer this question, SAP-affiliated authors facilitated three co-design workshops at a branch office in the US.
After providing some background, we first describe goals of each workshop and their process with some intermediate results respectively. This is followed by participant feedback and our recommendations on improving the hybrid work model and workplace from the perspectives of communication, change management, and organizational culture. In particular, although these workshops created an initial structure of hybrid schedules, software practitioners at this site still expected to regularly re-visit their hybrid work model and calibrate its arrangements as an outcome of their team retrospectives. 
We argue that our co-design adaptation, lessons learned according to this practice, and recommendations for a hybrid work model can provide valuable insights for other software organizations adapting to the ``new'' normal in the post-pandemic world.

\section{BACKGROUND}

Co-design refers to the collective creativity of both designers and non-designers working together in the design process~\cite{sanders2008co}. As a participatory approach to design, it strives to cater various stakeholders' fuzzy needs as well as foster consensus, creativity, and collaboration.
Human-Computer Interaction (HCI) and Software Engineering (SE) research has increasingly demonstrated interest in co-designing for marginalized populations and underrepresented groups. 
In SE, prior co-design practice also enabled developers to articulate imprecise requirements in software-intensive development~\cite{hehn2019integrating}, and hence improve user satisfaction.
While recent reports often indicate developers' diverse work preferences towards remote and on-site arrangements~\cite{ford2021tale}, co-design could be a suitable approach to address an organization's fuzzy needs during transitioning to a particular work model.

To leverage practitioners' creative potential, we incorporated this site's design thinking (DT) practice into the co-design process.
For this SAP site, DT has been applied as a collaborative and iterative approach that explores and elaborates the value proposition of a product in the early stages. Its DT approach aims to establish empathy for end users, and reduce the likelihood of costly adaptations~\cite{hildenbrand2012intertwining}.
Building upon DT's success with \textit{customer co-innovation}, i.e., an SAP activity that connects software developers and product managers with customers and articulates business and technical requirements of a product,
we anticipate improved co-design outcomes by leveraging participants' skills acquired from the site's reliable DT approach.

\section{CO-DESIGN WORKSHOPS}

\subsection{Site and Participants}
These workshops were organized at an SAP branch office at Newport Beach, California. SAP is a multinational corporation known for its integrated enterprise software. This branch office has started operating since 2019. This office practices Large-Scale Scrum (LeSS) to scale and optimize multi-team collaboration for software development~\cite{larman2016large}. Meanwhile, it has experienced large growth in organization size while working remotely during the pandemic. Until recently, many remotely onboarded software developers had not yet experienced this newly operated office. As a global software development organization, this site bears global dependencies, i.e., several major products here depend on internal services that are under development by teams at other locations. Starting 2022, this office has been actively preparing for hybrid work arrangements within the framework of a global corporate initiative called ``flex work''~\cite{tuan2022next}.

\begin{figure*}
    \includegraphics[width=\textwidth]{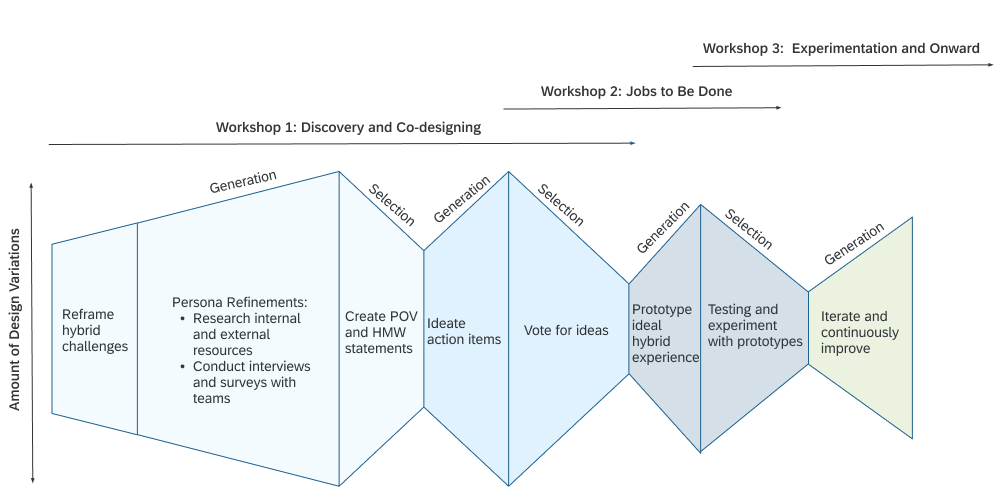}
    \caption{The overview of our co-design approach and the timeline of workshops. March 11th, 2022 was the initial \textit{Discovery and Co-design}. March 16th introduced \textit{Jobs-to-Be-Done} to specify the ideas produced in the first workshop. A month later on April 15th, we started the \textit{Experimental Design} for hybrid work.}\label{fig:framework}
\end{figure*}

Participants in the co-design workshops consisted of software engineers, UX designers, and product managers representing the major software development roles at the site. The number of participants varied from 17 to 22 across the three workshops with a common core of people.
The first workshop was organized in March 2022 including a portion of dedicated remote participants, and in the rest two they also participated in-person for improved workshop experience. The site's actual hybrid transition started in July 2022 (see Figure \ref{fig:framework}). 

All co-design artifacts such as sticky notes and whiteboard collaboration spaces were digitized for thematic analysis by the authors, and the emerging themes are reported in the following sections.
Additionally, we released agendas and methodological guidelines for these co-design workshops in a public repository\footnote{\label{foot:supply}Co-design hybrid: \url{https://github.com/co-design-hybrid/co-design-hybrid}}.

\subsection{Workshop 1: Discovery and Co-design}
\begin{figure}
\includegraphics[width=\columnwidth]{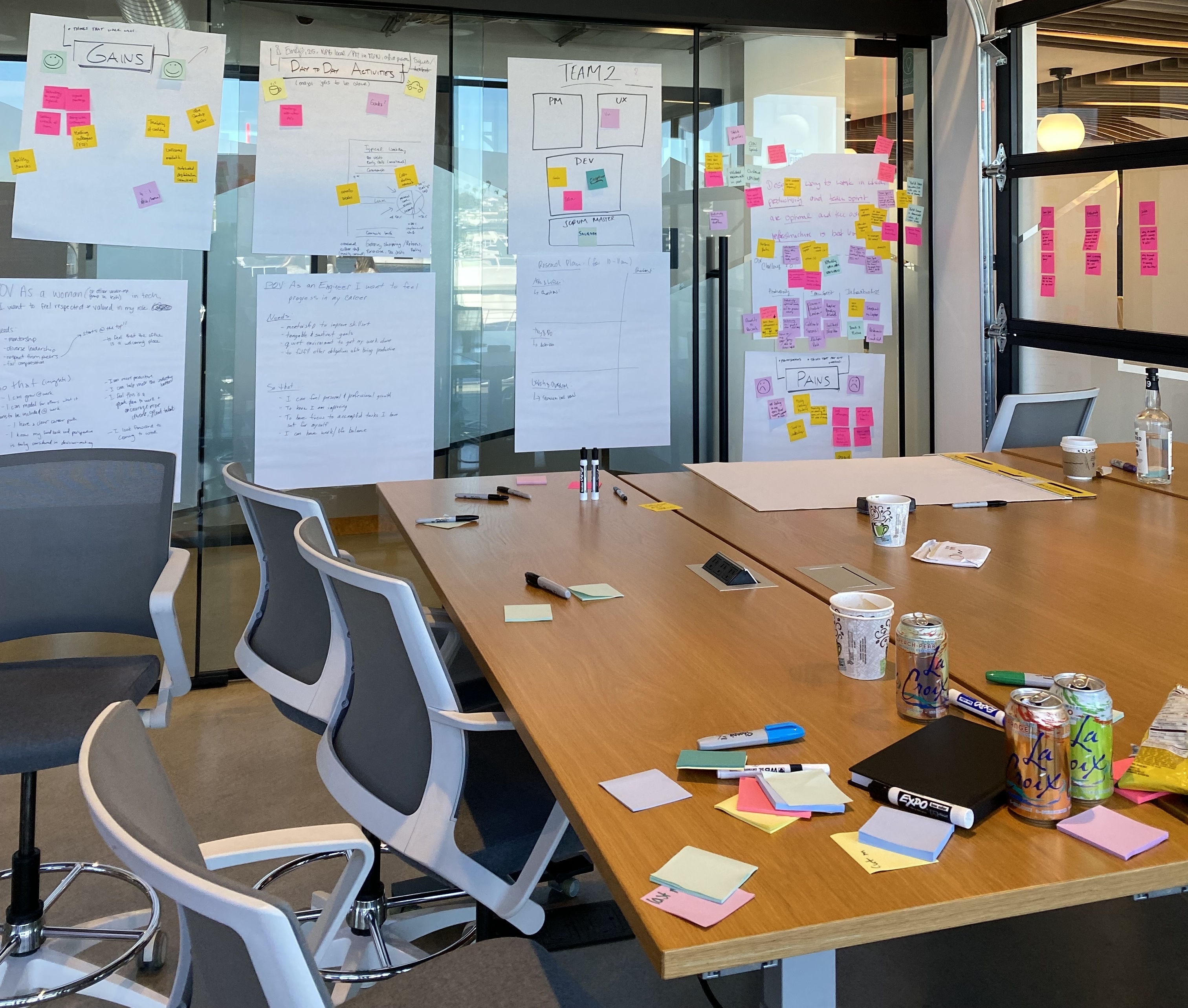}
\caption{Participants might add notes onto whiteboards for refining the personas, and identify core themes of a persona's workplace experience and needs.}\label{fig:room}
\end{figure}

The first workshop was a daylong event for participants to co-design the site's hybrid experience, and set up initial action items for pilot experiments and improvements. This workshop and the following were held in a large conference room with whiteboard spaces and a wall display (see Figure \ref{fig:room}). After a short introduction, participants were randomly assigned to four co-design teams.

\textit{Reframe Challenge and Scope.} The first step of this workshop was to understand, refine and reframe the underlying challenges for transitioning to hybrid at this site. 
We encouraged participants to raise questions about hybrid challenges, and record and paste these questions onto a collaborative space for team reviews. From this step, participants anchored the overall direction for how to address issues in hybrid work for this entire workshop.

\textit{Research and Synthesis.} The second step enabled participants to explore the needs of their target personas in a hybrid workplace. Personas have been widely used in design and DT projects as an approach to build empathy between users and designers~\cite{pruitt2010persona}. As a starter, facilitators provided the three main persona examples in the domain of engineering, UX, and product management, which were based on existing roles at the site.
Participants were encouraged to leverage a broad variety of curated research resources (e.g., internal technical reports, research literature, and press articles on hybrid work) to further elaborate on the underlying challenges of each given persona (see example personas in the public repository\textsuperscript{\ref{foot:supply}}). 

\textit{Ideate, Prototype, and Validate.} Finally, in DT terms, participants ``ideated'' on solutions and action items collaboratively to address the needs of their personas. As a prerequisite, each team first created several Point-Of-View (POV) statements for their target persona to address specific user needs and provide improvement insights, e.g., ``as a UX designer, \textit{I need} a collaborative open space to run customer workshops with my team \textit{so that} we can get maximum empathy and feedback from customers.'' Then each participant elaborated on their ideas with POV statements, and proceeded with a team discussion. They prototyped their top-team-rated ideas by creating a concrete plan that included a set of action items for the organization to experiment and improve its workplace. The prototype validation was further carried out with a team discussion and voting activity. At the end of this workshop, each co-design team provided a presentation of proposed improvements to the organization.

The first workshop enabled participants to brainstorm on explicit solutions to address these statements, e.g., proposing the organization arrange training and mentoring sessions for career development, and create interest groups for re-establishing a sense of belonging at work, while some participants were concerned with few action items strayed off-topic and general work discussion.
Co-design teams mainly voted for the ideas that were actionable for themselves, and also the ones perceived  as substantially leading to an improved workplace experience. 

\subsection{Workshop 2: Development Jobs-to-Be-Done}

The second workshop aimed at specifying a focused action plan for an ideal hybrid workplace experience. Since the first workshop created a diverse and overwhelming number of action items, the second workshop aligned design goals for a hybrid workplace experience around main development tasks. This alignment process was guided by a business strategy, \textit{Jobs-to-Be-Done}, often adopt at this site, which aims for articulating the practical needs of end users and customers~\cite{christensen_know_2016}. Thus, this workshop could create an initial structure of hybrid schedule, and set tracking that monitored these items' progress.

We first introduced a categorization of activities that are essential for overall software design and delivery.
This category of activities incorporated findings in the last workshop into the \textit{Collaboration Model}, the site's software development standards for product management and engineering to discover, design, and deliver products collaboratively. These categories included \textit{Development Preparation}, \textit{Heads-down Work}, \textit{Reviews \& Retrospectives}, \textit{Customer Co-innovation}, \textit{Team Building} and \textit{Personal Development}.
Participants placed their recurring development activities into these categories for an ideal workplace experience: distinguishing the activities that favor on-site from tasks that prefer remote and quiet environments.

To manage the overwhelming action items that emerged from the first workshop, the facilitators introduced a visual Kanban board where participants could create \textit{improvement} tickets for their action items. Moreover, all employees of this office had access to the board and thus could comment on and keep track of the organization's progress with potential improvements.

The second workshop created an initial hybrid schedule for this site, which facilitated most collaborative and mentoring activities on-site, and focused work remotely. 
When creating the hybrid schedule, participants realized that optimizing their work required the support of tools and on-site infrastructure. 
Additionally, the Kanban board urged the organization to progress with highly prioritized action items, e.g., providing office benefits for working on-site days.

\begin{figure*}
    \includegraphics[width=\textwidth]{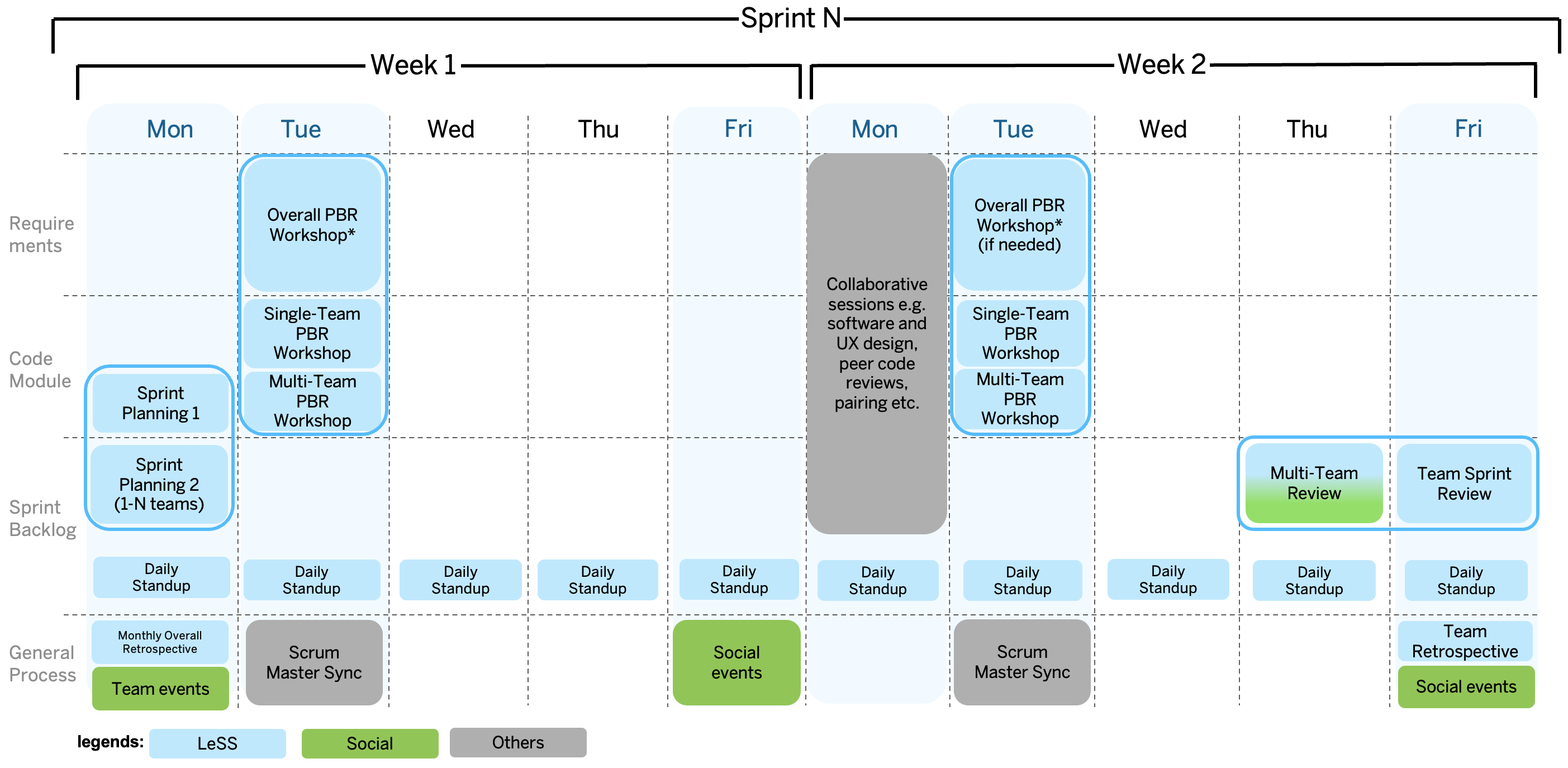}
    \caption{General guidelines of a two-week Sprint for collaborative activities at the site. Mondays, Tuesdays and Fridays are on-site days for collaborative events, and Wednesdays and Thursdays are off-site days for focused work. Daily standup meetings can either be online or in-person forms to align with the overall schedule.
    \\\hfill
    *PBR is short for Product Backlog Refinement meetings as described in~\cite{larman2016large}.}\label{fig:schedule}
\end{figure*}

\subsection{Workshop 3: Experimental Designs and Piloting}
The last workshop was a short session that designs small experimental improvements for challenging action items and solicited teams to experience hybrid schedules.

Facilitators organized participants to review the Kanban board together to focus on several long-lasting and challenging topics (e.g., interactive whiteboard or writable wall were not sufficient in conference rooms). The workshop delegated each topic to at least two participants. These small groups collaboratively plan and set weekly progress goals (e.g., contact facilitator management team, purchasing, and installation) in a pair working session.

In addition, a few development teams solicited by participants voluntarily experimented with working on a hybrid schedule.
Meanwhile, all participants continuously provided feedback on the advances during weekly check-ins on-site and on the transparent Kanban board, e.g., concerns were raised about allocating on-site review meetings on Fridays overlapping with most team-building activities.

\subsection{Concluding Interviews}

We conducted semi-structured concluding interviews with available participants about their experiences with workshops and ensuing hybrid work, after two weeks (one Sprint) of an office-wide hybrid schedule.

Participants perceived being a part of the workshops as an interactive, enjoyable, and engaging experience. They enjoyed attending these creative events in-person for timely confirmation and feedback, and also fostered the collective creativity of the participant group. With the last two workshops, they formed a self-organized group with a shared mission to transparently address critical issues in their hybrid work. Their experience during this collaborative process was perceived as a natural way to leverage organizational resources, e.g., design thinking practice within SAP, existing site infrastructure, and working as a cross-functional group on-site. 

While overwhelming personal needs and concerns were raised, some participants sensed a disconnect between the workshop results and the ensuing hybrid practice. For instance, predefined on-site days needed to be agreed upon for achieving high-quality team synchronization, and promoting highly collaborative work such as Product Backlog Refinement meetings. Additionally, some challenging action items took an extended time to progress, e.g., although participants preferred that product teams could collaborate with their product neighborhoods in co-located sessions, compromises had to be made under globally distributed settings.

\section{HYBRID WORKPLACE EXPERIENCE}

Based on the workshop outcome, we initialized a hybrid schedule that organized recurring software development tasks such as common Scrum ceremonies into on- and off-site schedules (see Figure \ref{fig:schedule}). 
For a typical five-day week during a Sprint, this site practiced three aligned on-site days for Scrum ceremonies of the basic Scrum and LeSS framework as well as architecture and UX design sessions.
These on-site days were primarily reserved for collaboration-intense sessions in-person, and some optional and informal team-building events. 
The remaining two days were remote days for focused standalone work sessions in-between; one was the No-Meeting-Wednesday to reduce interruptions.
The rationale behind this formation was to fit Scrum activities in a 2-week time frame: started with planning in week 1, and ended with reviews and retrospectives in week 2. 
After all, developers might utilize their on-site workplaces and assigned desks at any time they prefer.

\subsection{Bridging Distance between On-Site and Remote Teams}
Communication and collaboration are key dimensions of productivity in team-based software development~\cite{forsgren2021space}.
Impaired communication and collaboration often exert escalating effects on other productivity dimensions, which result in re-work, distrust, and dissatisfaction. 
For this distributed software development organization, hybrid meetings--i.e. meetings whose attendees comprise on-site and remotely off-site--are bound to happen. 
According to our participants, most of these meetings were unpleasant experiences due to stressful meeting times caused by timezone differences, imbalanced awareness and common ground among attendees who did not share physical meeting space, and skepticism towards off-site attendees' commitment.

Our primary suggestion is to coordinate and leverage on-site time deliberately while limiting the number and frequency of hybrid meetings as such.
Prior lessons learned from this site suggested that the growing distrusts from avoidable hybrid meetings substantially harmed team spirit and work morale, especially for inexperienced or newly formed teams. 
Besides, hosting unnecessary hybrid meetings could defeat the purpose of on-site hours for in-person collaboration and fostering social interaction, as mentioned by several in the concluding interviews.

For unavoidable hybrid meetings with off-site attendees, available and well-equipped conferencing devices, a software toolset for hybrid collaboration, and an inclusive meeting format are fundamentals. Based on participants' feedback, sufficient phone booths and access to various sizes of well-equipped conference rooms are not only essential for the meeting experience, but also for creating a focused and distraction-free open workplace. 
For sharing design artifacts in hybrid meetings, conference rooms with interactive whiteboards (e.g. SurfaceHub and Jamboard) substantially help with cross-organization and -function collaboration. Moreover, providing a complete set of collaboration tools is equivalently essential as technical engineering systems~\cite{jackson2022collaboration}. 
This site supplies each developer with a suite of hybrid collaboration tools, including Mural, Teams, Slack, LucidChart, etc., to facilitate collaborations with remote colleagues. 
Finally, experience from the first workshop suggested that hybrid meetings could apply an inclusive meeting format such as Roundtable during discussion. 
Such an inclusive format ensures remote participants have equal opportunities to speak up.

\subsection{Building Cornerstones of a Seamless Transition}
According to participant feedback from the hybrid experiment, the most essential personal adjustments were allocating commute and core work hours into in-person days. 
With the wave of fully remote work during COVID-19, the work-from-home schedule provided workers with small breaks and personal time throughout the workday.
We provide two major suggestions on initiating a seamless hybrid transition and preserving the flexibility since working from home: one, providing local flexibility of working hours and location at the team level; and two, incentivizing employees to experience hybrid arrangements earlier.

First, adjustments on individual and team schedules are hardly completed overnight, and decision-makers may consider providing some room for flexibility. According to the concerns from participants, strict regulation on core hours increased their stress over time management in addition to the reluctance of commuting. Thus, when they attended meetings or were on-call beyond regular core hours, providing flexibility on working hours and location would reduce their stress, and safeguard trust within the team. For instance, attending an early morning meeting with Europe teams at home and then commuting to the office later at noon should not raise skepticism towards their work ethics. 

Second, participants with prior hybrid or on-site experience, e.g., participation in hybrid experimentation, or working in an office before the pandemic
often led to more positive feedback with the office-wide hybrid schedule. Socializing with teammates in-person, incrementally adjusting to individual and local teams' hybrid schedules, and being familiar with office infrastructure and benefits (especially for remotely onboard members) substantially improved participants' workplace experience. Besides monetary incentives, there were many other events for motivating employees to experience the hybrid workplace, e.g., customer site visits, social outings with teammates at or near the office, interest group events with academic collaborators and colleagues, and office-sponsored lunch events. Being physically together during these events, cross-function groups could establish common grounds and foster personal trust for efficient collaborations~\cite{olson2000distance, moe2008understanding}.

\subsection{Fostering Culture and Trust for Collaboration and Deep Work}

High-performing teams have strong foundations in trust~\cite{tuckman1977stages}. 
Trust is a combination of personal connection, consistency, and experience.
Trust is also a key element of efficient communication, exerting influence on organizations and design.
SAP has had a large success in integrating end-to-end business processes that cross traditional internal departmental silos. As such, it has gained a competitive advantage by establishing a culture of high trust and collaboration. This is especially true for a new team and new product development of any sufficient complexity where the interfaces between teams and systems are unclear. Allowing teams to co-work and form together in a modern workplace drives the desired function of rapid learning and trust building and accelerating the development of integrated systems and components.

While minimizing the distance between engineers, UX designers and product managers is important from an information flow perspective, any sufficiently complex work to be solved also needs time to focus on individual tasks for longer periods of time; sometimes referred to as \textit{deep work}, i.e., professional activities performed in a state of distraction-free concentration~\cite{newport2016deep}.
Balancing highly collaborative work with deep work is consistent with the site's core business value around building enterprise-grade products that leverage systems thinking and decision support planning systems efficiently.
Therefore, hybrid, from the view of mixing highly collaborative and close proximity work with deep work is substantially ingrained in this organization's culture, and a large part of its customer and financial success. 
Thus a ``dual" style of working is naturally the mode of most enterprise software engineers, but perhaps rather a more delineated work pattern in complex sub-system programming and integration as required for enterprise software.

With COVID-19 shifting SAP teams and thus communication patterns to a completely distributed model, a shift to more asynchronous and remote communication methods also occurred.
It is therefore important for management to re-establish existing and re-think new complimentary working principles along people, process, and technology pillars, i.e., balancing healthy team characteristics, organizational productivity, and individual flexibility while ensuring equal career advancement opportunities for off- and on-site members.
This will be an area of interest and research for SAP and likely other large-scale software firms in the coming years.

\subsection{Practicing Co-design to Articulate Specifics in Hybrid Workplace}

To locally optimize the hybrid workplace experience and articulate the specific needs of practitioners at other similar large software development organizations, we recommend applying this co-design format\textsuperscript{\ref{foot:supply}} on local teams with the following considerations.
First, participant selection and persona example illustration of the workshop should be representative based on the site's team roles and demographics; second, design formats and resources should align with the local team's daily practice to achieve optimal outcomes of co-design activities; and third, intermediate co-design results should be communicated to organization members outside the workshops and solicit feedback, in our case, through a Kanban board.

Moreover, practitioners may consider planning and adapting this format for other types of organizations through some rigorous structure such as the GQM approach~\cite{caldiera1994goal}, especially when there were doubts about the worthiness of a hybrid transition. 
Particularly, the specific process could be developed based on the goals of the hybrid transition (e.g., to improve collaborative engineering and creative work), questions about what can be improved for transitioning process and beyond (e.g., site infrastructure and career development opportunities), and metrics to evaluate whether the transition was satisfying (e.g., unplanned meeting amount and frequency, and employee self-reported satisfaction).

\section{CONCLUDING REMARKS}
In this article, we provide a novel approach of co-designing for a hybrid workplace experience in software development, and present our takeaways for improving such a hybrid arrangement continuously. Co-design workshop formats can also help other organizations create their own paradigm of hybrid work via a bottom-up approach. Following a continuous improvement process, we will regularly re-visit our hybrid arrangement to calibrate organizational preferences and the site's collaborative development practice.

\section{ACKNOWLEDGEMENT}
We would like to thank all anonymous reviewers. Chou, Wang and Redmiles are supported in part by UC Irvine, Donald Bren School of ICS. Prikladnicki is partially funded by Fapergs and CNPq, Brazil.

\bibliographystyle{IEEEtran}
\bibliography{softwareSI.bib}

\begin{IEEEbiography}{Zhendong Wang}{\,} is currently working towards his Ph.D. degree in Software Engineering at the University of California, Irvine.  His research focuses on leveraging and supporting developer expertise in distributed software development. Wang received an M.S. degree in Software Engineering from the University of California, Irvine in 2018. Contact him at zhendow@uci.edu.
\end{IEEEbiography}

\begin{IEEEbiography}{Yi-Hung Chou}{\,} is currently a software development/research intern at SAP Newport Beach. He is pursuing M.S. Software Engineering at University of California, Irvine. His research interest is at the intersection of human-computer interaction and software engineering. Contact him at yihungc1@uci.edu.
\end{IEEEbiography}

\begin{IEEEbiography}{Kayla Fathi}{\,} is a user experience designer at SAP, working within an innovation unit, Industry Cloud. Fathi received her B.S. in Cognitive Science from the University of California, Berkeley. Contact her at kaylafathi@gmail.com.
\end{IEEEbiography}

\begin{IEEEbiography}{Tobias Schimmer}{\,} is a software engineering practitioner with SAP, and an adjunct professor at the University of Mannheim. He is responsible for SAP's Industry Cloud collaboration model and respective product engineering operations. His research focuses on multi-team collaboration, empirical software engineering, and data-driven process improvements. Schimmer received his Ph.D. degree from the University of Mannheim in 2008 and has been a visiting researcher with University of California, Irvine, since 2006. Contact him at tobias.schimmer@sap.com.
\end{IEEEbiography}

\begin{IEEEbiography}{Peter Colligan}{\,} is a senior executive manager and head of engineering for SAP's Industry Cloud. Contact him at peter.colligan@sap.com.
\end{IEEEbiography}

\begin{IEEEbiography}{David Redmiles}{\,} is a Professor in the Department of Informatics at the University of California, Irvine, in the Donald Bren School of Information and Computer Sciences. Redmiles received his Ph.D. in Computer Science in 1992 from the University of Colorado, Boulder. His research integrates the areas of software engineering, human-computer interaction, and computer-supported cooperative work. Over the years, his research group has investigated themes of cognitive support for software developers; issues of trust and emotion affecting software teams; behaviors of participants in social software development platforms; global software engineering; end-user software development; and more. He is a member of the ACM and the IEEE Computer Society. He was designated an ACM Distinguished Scientist in 2011 and a Fellow of Automated Software Engineering in 2009. Contact him at redmiles@ics.uci.edu.
\end{IEEEbiography}

\begin{IEEEbiography}{Rafael Prikladnicki}{\,} is an associate professor in the School of Technology at the Pontifical Catholic University of Rio Grande do
Sul, Porto Alegre, 90450171, Brazil. Contact him at https://www.pucrs.br/researchers/rafael-prikladnicki/ or rafaelp@pucrs.br.
\end{IEEEbiography}

\end{document}